\documentclass[aps,prd,twocolumn,superscriptaddress,nofootinbib,amsmath,amssymb,floatfix]{revtex4-2}
\usepackage{multirow}
\usepackage[T1]{fontenc}
\usepackage{lmodern}
\usepackage{graphicx}
\usepackage{xcolor}
\usepackage{bm}
\usepackage{booktabs}
\usepackage{hyperref}
\usepackage{microtype}

\hypersetup{
  colorlinks=true,
  citecolor=blue!55!black,
  linkcolor=red!55!black,
  urlcolor=blue!65!black
}
\graphicspath{{../code/plots/}}
\newcommand{\mchi}{m_\chi}
\newcommand{\mphi}{m_\phi}
\newcommand{\sigmabar}{\overline{\sigma}_{\chi e}}
\newcommand{\dd}{\mathrm{d}}

\begin{document}

\title{Direct Detection of Light Self-Interacting Dark Matter via Electronic Collective Excitations}
\author{Wen-Na Yang}
\email{wennayang@njnu.edu.cn}
\affiliation{Department of Physics and Institute of Theoretical Physics, Nanjing Normal University, Nanjing, 210023, China}

\author{Wu-Long Xu}
\email{wlxu@cdtu.edu.cn}
\affiliation{School of Integrated Circuits, Chengdu Technological University, Chengdu 611730, P. R. China}

\author{Wenyu Wang}
\email{ wywang@bjut.edu.cn}
\affiliation{School of Physics and Optoelectronic Engineering,
Beijing University of Technology, 100124, Beijing, China}

\author{Ning Liu}
\email{liuning@njnu.edu.cn}
\affiliation{Department of Physics and Institute of Theoretical Physics, Nanjing Normal University, Nanjing, 210023, China}
\affiliation{Nanjing Key Laboratory of Particle Physics and Astrophysics, Nanjing, 210023, China}

\begin{abstract}
% Models of light dark matter naturally invoke a light mediator to facilitate interactions with the Standard Model. If sufficiently light, this mediator induces long-range self-interactions among dark matter particles, offering a compelling resolution to small-scale structure anomalies. However, direct detection of light self-interacting dark matter (SIDM) is challenging for conventional detectors. In this work, we study the sensitivity of searching for the light SIDM accelerated by high-energy cosmic rays in the Silicon detector. Leveraging the electronic collective excitations, we derive 90\% C.L. exclusion limits using public SENSEI and DAMIC ionization data. Our constraints can cover a portion of the light SIDM parameter space preferred by galactic small-scale anomalies.

Models of light dark matter often invoke a light mediator to facilitate interactions with the Standard Model. If sufficiently light, this mediator can induce long-range self-interactions among dark matter particles, offering a compelling  resolution to small-scale structure anomalies. However, direct detection of light self-interacting dark matter (SIDM) remains challenging for conventional detectors. 
%\textcolor{red}{as the energy deposited by nonrelativistic halo DM is typically below experimental thresholds}.
In this work, 
we investigate the sensitivity of searches for light SIDM accelerated by high-energy cosmic rays in silicon detectors. Leveraging the electronic collective excitations, we derive 90\% C.L. exclusion limits using public SENSEI and DAMIC-M ionization data. 
Our constraints can cover a portion of the light SIDM parameter space favored by galactic small-scale anomalies.
% Self-interacting dark matter (SIDM) provides a well-motivated framework for addressing small-scale structure anomalies in the standard cosmological model, while probing light SIDM in laboratory experiments remains challenging because conventional direct detection searches rapidly lose sensitivity at low DM masses. \textcolor{red}{In SIDM scenarios in which the same light scalar mediates DM self-interactions and couples to electrons, Galactic cosmic-ray electrons can upscatter halo DM into an energetic component accessible to low-threshold semiconductor detectors. In this work, we explore this complementary search channel in an SIDM framework with a light scalar mediator coupled to electrons, focusing on CRDM-induced electronic excitations and the plasmon-enhanced response of silicon.} Using low-threshold silicon data from the SENSEI \textcolor{red}{DAMIC} experiment, we derive 90\% confidence-level constraints in the $(m_\chi,m_\phi)$ plane for fixed benchmark values of the reference DM-electron scattering cross section $\bar{\sigma}_{\chi e}$. \textcolor{red}{We find that the combination of cosmic-ray upscattering and collective plasmon excitations in silicon can probe SIDM-motivated parameter space in the keV-GeV DM mass range. This demonstrates the potential of low-threshold semiconductor detectors as a complementary direct detection  probe of light SIDM.}

\end{abstract}

\maketitle

\section{Introduction}
Dark matter (DM) constitutes the dominant matter component of the Universe~\cite{Planck:2018vyg}, yet its nature remains unknown~\cite{Bertone:2004pz}. The standard $\Lambda$CDM framework, in which DM is assumed to be cold and collisionless, accounts for a broad range of cosmological observations, including cosmic microwave background anisotropies and the large-scale distribution of matter~\cite{Planck:2018vyg}. However, on galactic and sub-galactic scales, collisionless cold DM faces several well-known challenges, such as the core-cusp problem~\cite{Flores:1994gz,Burkert:1995yz,Salucci:2007tm,deBlok:2009sp,
Walker:2011zu,Oh:2010mc}, the too-big-to-fail problem~\cite{Boylan-Kolchin:2011qkt,
BoylanKolchin:2012qa}, and the diversity of galactic rotation curves~\cite{Oman:2015xda,Bullock:2017xww}. These small-scale structure anomalies have motivated self-interacting dark matter (SIDM) scenarios~\cite{Spergel:1999mh,Vogelsberger:2012ku,Rocha:2012jg,Tulin:2017ara}, with a phenomenologically favored transfer cross section per unit DM mass of $\sigma_T/\mchi\simeq 0.1$--$10~\mathrm{cm^2/g}$ on galactic scales~\cite{Peter:2012jh,Zavala:2012us,Elbert:2014bma,
Kaplinghat:2015aga}. Such self-interactions can arise naturally from the exchange of a light mediator. If sufficiently light, the mediator induces long-range interactions among DM particles, leading to a velocity-dependent transfer cross section
across different halo scales
~\cite{Loeb:2010gj,Tulin:2013teo,Colquhoun:2020adl}.

A light mediator can also facilitate interactions with Standard Model particles, providing a laboratory avenue for probing SIDM~\cite{Qin:2011za,DelNobile:2015uua,Wang:2022akn,PandaX-II:2021lap,Xu:2024iny}. However, direct detection of light SIDM remains challenging for conventional detectors, as the energy deposited by non-relativistic halo DM is typically below experimental thresholds. 
Low-threshold noble-liquid and semiconductor experiments have recently extended direct detection sensitivity to sub-GeV DM masses~\cite{Angle:2011th,DarkSide:2018bpj,SuperCDMS:2018mne,DAMIC:2019dcn,XENON:2019gfn}. For a mediator coupled to electrons, semiconductor targets offer a particularly sensitive channel, as eV-scale electronic excitations can generate measurable ionization signals via DM-electron scattering~\cite{Essig:2011nj,Essig:2012yx,Essig:2015cda}. In particular, silicon Skipper CCDs achieve single-electron resolution, enabling searches for single- and few-electron ionization events~\cite{Tiffenberg:2017aac,SENSEI:2018dpa,SENSEI:2020dpa,DAMIC-M:2023gxo}. Nevertheless, the energy and momentum transfers available from non-relativistic halo SIDM remain kinematically restricted.

Cosmic ray accelerated DM provides a complementary avenue for probing light SIDM beyond this kinematic limitation. High-energy galactic cosmic-ray electrons can upscatter halo DM to (semi-)relativistic velocities~\cite{Bringmann:2018cvk,Cappiello:2018hsu,Ema:2018bih,Dent:2020syp,Xia:2020apm,Wang:2021nbf,Su:2022wpj,PandaX:2024pme}, thereby allowing the accelerated SIDM to access the energy--momentum transfer regime where electronic collective excitations in silicon dominate. The electronic response of silicon is characterized by the energy-loss function $\mathrm{Im}[-\epsilon^{-1}(q,\omega)]$, which exhibits a pronounced plasmon resonance at $\mathcal{O}(10)~\mathrm{eV}$~\cite{Knapen:2021run,Hochberg:2021pkt,Essig:2024ebk,Dreyer:2026bmz}. The plasmon resonance enhances the DM-electron scattering rate in silicon, producing an increased few-electron ionization signal, as demonstrated in previous studies~\cite{Liang:2024xcx,Guo:2024sqh,Gong:2026dte}.

In this work, we investigate the sensitivity of silicon detectors to light SIDM accelerated by high-energy cosmic ray electrons. 
We consider a light scalar mediator coupled to both DM and electrons, which mediates DM-electron scattering.  
Using public SENSEI and DAMIC-M ionization data, we derive 90\% C.L. exclusion limits in the $(m_\chi,m_\phi)$ plane for benchmark values of the reference DM-electron scattering cross section $\bar{\sigma}_{\chi e}$. 
Our constraints can cover a portion of the light SIDM parameter space favored by galactic small-scale anomalies, demonstrating that low-threshold silicon detectors provide a complementary probe of light SIDM. 
This paper is organized as follows. Section~II describes the calculation of DM self-interactions, Sec.~III presents the cosmic ray accelerated SIDM flux and the signal rate from electronic collective excitations in silicon, Sec.~IV gives the resulting constraints, and Sec.~V summarizes our conclusions.

\section{Dark-matter self-interaction}
We consider a phenomenological framework in which non-relativistic DM scattering is described by an attractive Yukawa potential:
\begin{equation}
    V(r)=-\frac{\alpha_\chi}{r}e^{-m_\phi r},
    \qquad
    \alpha_\chi=\frac{g_\chi^2}{4\pi}.
    \label{eq:yukawa}
\end{equation}
This potential arises from the exchange of a scalar mediator $\phi$, with the interaction Lagrangian
\begin{equation}
    \mathcal{L}_{\rm int}
    =
    g_\chi\phi\bar{\chi}\chi,
    \label{eq:dark-interaction}
\end{equation}
where $g_\chi$ is the dark-sector coupling. Since forward scattering transfers little momentum, the efficiency of momentum exchange in DM halos is characterized by the momentum-transfer cross section
\begin{equation}
 \sigma_T
 =
 \int d\Omega\,
 (1-\cos\theta)\frac{d\sigma}{d\Omega}.
 \label{eq:sigmaT}
\end{equation}
% The angular weight suppresses small-angle scattering events and measures the efficiency with which scattering redistributes momentum.
The scattering regimes of the Yukawa potential are characterized by the
dimensionless parameters
\begin{equation}
 \kappa=\frac{\mchi v}{2\mphi},
 \qquad
 \beta=\frac{2\alpha_\chi\mphi}{\mchi v^2},
 \label{eq:kappabeta}
\end{equation}
where $v$ is the relative velocity. The parameter $\kappa$ is the dimensionless momentum in the center-of-mass frame and is defined as the ratio of the interaction range to the de Broglie wavelength. The parameter $\beta$ measures the ratio of the potential energy to the kinetic energy. 
The Born limit corresponds to the condition $2\beta\kappa^2\ll1$, while the quantum and semiclassical regimes are
distinguished by $\kappa\lesssim1$ and $\kappa\gtrsim1$,
respectively~\cite{Tulin:2013teo,Colquhoun:2020adl}. The explicit expressions for $\sigma_T$ used in our numerical evaluation are provided in Appendix~\ref{app:selfinteraction}.

To account for the distribution of relative velocities within a halo, we
compute the velocity-averaged transfer cross section
\begin{equation}
 \langle\sigma_T\rangle=
 \int_0^\infty\dd v\,f(v)\sigma_T(v),
 \label{eq:velocityaverage}
\end{equation}
using the Maxwell-Boltzmann distribution
\begin{equation}
 f(v)=\frac{32v^2}{\pi^2v_0^3}
 \exp\!\left[-\frac{4v^2}{\pi v_0^2}\right]\,.
 \label{eq:velocitydistribution}
\end{equation}
In this work, we adopt a characteristic velocity of the Milky Way, $v_0 = 200~\mathrm{km\,s^{-1}}$. The SIDM target region is defined by the condition
\begin{equation}
0.1 < \frac{\langle\sigma_T\rangle}{m_\chi} < 10~\mathrm{cm^2/g},
\label{eq:sidmtarget}
\end{equation}
which defines the SIDM-favored parameter space considered in our analysis.

\section{Direct Detection of Cosmic Ray Accelerated Light SIDM}

\subsection{Flux of Cosmic Ray Accelerated SIDM}
Galactic cosmic ray electrons can upscatter non-relativistic halo DM
into an energetic component. We consider a Dirac fermion $\chi$ coupled to
electrons through a light scalar mediator $\phi$, with the interaction
Lagrangian
\begin{equation}
 {\cal L}_{\rm int}=
 g_\chi\phi\bar\chi\chi+g_e\phi\bar e e,
 \label{eq:lagrangian}
\end{equation}
where $g_\chi$ and $g_e$ denote the mediator couplings to DM and electrons,
respectively.

The differential accelerated SIDM flux at Earth is obtained by convolving the
cosmic-ray electron spectrum with the DM-electron scattering cross section,
\begin{equation}
 \frac{\dd\Phi_\chi}{\dd T_\chi}=
 D_{\rm eff}\frac{\rho_\chi}{\mchi}
 \int_{T_e^{\rm min}}^\infty\dd T_e\,
 \frac{\dd\Phi_e}{\dd T_e}
 \frac{\dd\sigma_{\chi e}}{\dd T_\chi},
 \label{eq:flux}
\end{equation}
where $T_\chi$ and $T_e$ denote the kinetic energies of the recoiling DM particle and the incident cosmic-ray electron, respectively. In the DM rest frame, the maximum kinematically allowed recoil energy is
$
 T_\chi^{\rm max}=
 \frac{2\mchi(T_e^2+2m_eT_e)}
 {2\mchi T_e+(\mchi+m_e)^2}
$.
Inverting this relation yields the minimum incident electron energy required to produce a DM recoil energy $T_\chi$:
\begin{equation}
T_e^{\min}
=
\left(\frac{T_\chi}{2}-m_e\right)
\left[
1 \pm
\sqrt{
1+
\frac{2T_\chi (m_e+m_\chi)^2}
{m_\chi (2m_e-T_\chi)^2}
}
\right]\,,
\end{equation}
where the signs $\pm$ correspond to $T_\chi>2m_e$ and $T_\chi<2m_e$. 
We take $\rho_\chi^{\rm local}=0.4~\mathrm{GeV\,cm^{-3}}$ and an NFW profile with $r_s=20~\mathrm{kpc}$. Assuming a spatially homogeneous CR distribution and integrating to $10~\mathrm{kpc}$ gives $D_{\rm eff}=8.02~\mathrm{kpc}$~\cite{Bringmann:2018cvk}. 
The CR electron flux is taken from the local interstellar spectrum (LIS),
$d\Phi_e^{\rm LIS}/dT_e=4\pi\,dI_e/dT_e$,
where $dI_e/dT_e$ is the differential intensity from Ref.~\cite{Boschini:2018zdv,Fang:2020dmi}.
The differential DM--electron scattering cross section is given by
\begin{equation}
\begin{split}
 \frac{\dd\sigma_{\chi e}}{\dd T_\chi}
 ={}&\frac{\sigmabar |F_{\rm DM}(q)|^2}{16\mu_{\chi e}^2}
 \frac{(2\mchi+T_\chi)(4m_e^2+2\mchi T_\chi)}
 {T_e^2+2m_eT_e}\,,
\end{split}
 \label{eq:dsigma}
\end{equation}
where $\mu_{\chi e}={\mchi m_e}/(\mchi+m_e)$ is the DM-electron reduced mass. The reference cross section is defined as $\sigmabar={\mu_{\chi e}^2g_\chi^2g_e^2}/
 {\pi(\mphi^2+q_0^2)^2}$, and the DM form factor is
\begin{equation}
 F_{\rm DM}(q)=\frac{\mphi^2+q_0^2}{\mphi^2+q^2}\,,
 \label{eq:fdm}
\end{equation}
where $q_0=\alpha m_e\simeq 3.7~\mathrm{keV}$ is the reference momentum,
$q^2=2m_\chi T_\chi$ is the squared four-momentum transfer, and $\alpha$ is
the fine-structure constant.
\begin{figure}[t]
 \centering
 \includegraphics[width=\columnwidth]{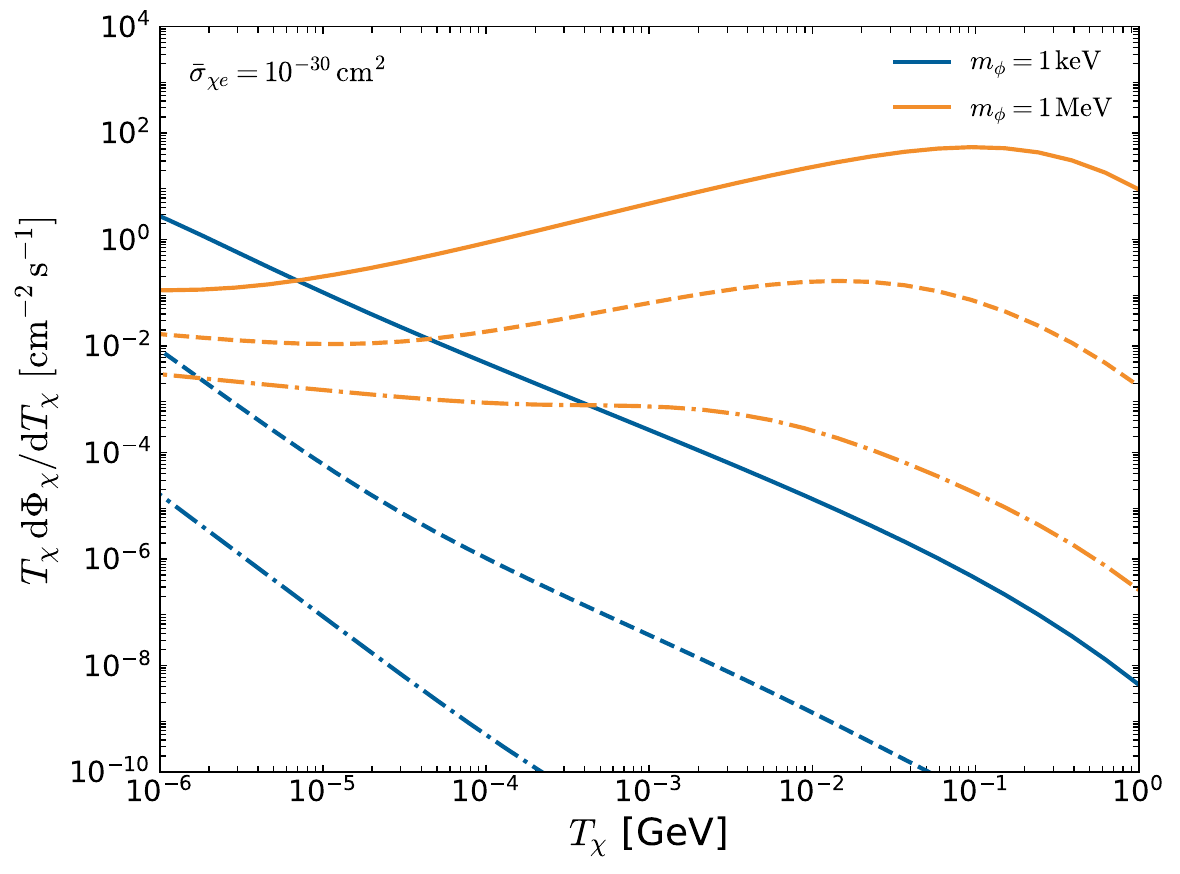}
 \caption{SIDM flux as a function of the DM kinetic energy $T_{\chi}$ from DM scattering with cosmic-ray electrons through a light scalar mediator with mass $m_{\phi}=1~\mathrm{keV}$ (blue lines) or $m_{\phi}=1~\mathrm{MeV}$ (orange lines). The reference DM-electron scattering cross section is fixed to $\bar{\sigma}_{\chi e}=10^{-30}~\mathrm{cm^{2}}$. The different curves correspond to different DM masses: $m_{\chi}=1~\mathrm{keV}$ (solid line), $m_{\chi}=10~\mathrm{keV}$ (dashed line), and $m_{\chi}=100~\mathrm{keV}$ (dash-dotted line).}
 \label{fig:flux}
\end{figure}

In Fig.~\ref{fig:flux}, we present the SIDM flux from upscattering of halo DM by galactic cosmic ray electrons for DM masses $m_\chi = 1~\mathrm{keV}$ (solid), $10~\mathrm{keV}$ (dashed), and $100~\mathrm{keV}$ (dash-dotted), mediator masses $m_\phi = 1~\mathrm{keV}$ (blue) and $1~\mathrm{MeV}$ (orange), and a reference cross section $\sigmabar = 10^{-30}~\mathrm{cm^2}$. 
The SIDM flux exhibits a stronger dependence on the DM mass for $m_\phi=1~\mathrm{keV}$ than for $m_\phi=1~\mathrm{MeV}$. This difference originates from the momentum dependence of the mediator form factor. For $m_\phi=1~\mathrm{keV}$, 
the form factor decreases appreciably over the momentum transfer range relevant to the production of the accelerated flux. Since $q^2=2m_\chi T_\chi$, increasing $m_\chi$ at fixed $T_\chi$ corresponds to a larger momentum transfer and hence to a stronger form-factor suppression. As the momentum transfer exceeds the mediator mass, the form factor approaches the scaling $|F_{\rm DM}(q)|^2\propto q^{-4}\propto (m_\chi^2T_\chi^2)^{-1}$. This introduces an additional suppression with increasing $m_\chi$. By contrast, for $m_\phi=1~\mathrm{MeV}$, the form factor remains nearly constant over much of the relevant kinematic range and therefore introduces only a weak additional dependence on $m_\chi$. As expected, cosmic-ray upscattering accelerates the DM particles to semi-relativistic velocities. In this work, we focus on upscattering by cosmic-ray electrons, since for sub-MeV DM the proton-induced flux is much smaller owing to the less efficient energy transfer from cosmic-ray protons to light DM.

\subsection{Electronic Excitations and Charge-Binned Ionization Signals in Silicon}
A semi-relativistic SIDM particle with sufficient kinetic energy can excite plasmons through DM--electron scattering in silicon detector.
The corresponding differential rate as a function of the deposited energy $\omega$ can be written as~\cite{Dreyer:2026bmz}
\begin{align}
 \frac{\dd R}{\dd\omega}&=
 \int\dd T_\chi\int\frac{d\Omega}{4\pi}
 \frac{\dd\Phi_\chi}{\dd T_\chi}
 \frac{\dd\sigma}{\dd\omega}\nonumber\\
 &= \int\frac{\dd T_\chi}{\rho_T}\int\frac{d\Omega}{4\pi}
 \frac{\dd\Phi_\chi}{\dd T_\chi}\left(\frac{E_\chi}{p_\chi}\right)
 \frac{\dd\Gamma}{\dd\omega}
 \label{eq:eventrate}\,,
\end{align}
where $\rho_T$ is the mass density of the detector, and $\Gamma$ is the transition rate for a given DM momentum $p_\chi$ in a detector
volume. The differential cross section for DM scattering off the electron density in a solid-state material with total mass $m_T$ is given by~\cite{Dreyer:2026bmz}
\begin{align}
\frac{\dd\sigma}{\dd\omega}
% =&
% \frac{m_T(g_eg_\chi)^2}{4\rho_T E_\chi v}
% \int\frac{\dd^3\mathbf Q}{(2\pi)^3}
% \frac{4\mchi^2-\omega^2+Q^2}
% {E'_\chi(Q^2+\mphi^2-\omega^2)^2}
% \nonumber\\
% &\times\frac{Q^2}{2\pi\alpha}
% \operatorname{Im}\!\left[-\frac{1}{\epsilon(\mathbf Q,\omega)}\right]
% \delta(E'_\chi-E_\chi+\omega).\\
&=\frac{m_T\sigmabar \pi}{4\mu_{\chi e}^2\rho_T E_\chi v}
\int\frac{\dd^3\mathbf Q}{(2\pi)^3}
{(4\mchi^2-\omega^2+Q^2)}|F_{\rm DM}(q)|^2
\nonumber\\
&\times\frac{Q^2}{2\pi\alpha}
\operatorname{Im}\!\left[-\frac{1}{\epsilon(\mathbf Q,\omega)}\right]
\delta(E'_\chi-E_\chi+\omega).
\label{eq:materialcrosssection}
\end{align}
Here $\mathbf Q$ is the momentum transferred to the material, $E_\chi$ ($E'_\chi$) is the initial (final) dark-matter energy, and $\epsilon(\mathbf Q,\omega)$ is the dielectric function. The DM form factor is defined in Eq.~\ref{eq:fdm}, with $q^2=Q^2-\omega^2$ for the scattering process in the detector.
The electronic response of silicon is characterized by the energy-loss function $\operatorname{Im}[-\epsilon^{-1}(\mathbf Q,\omega)]$, which we compute using the \textsc{DarkELF} package~\cite{Essig:2015cda,Hochberg:2021pkt,Knapen:2021bwg,Essig:2024ebk}. This package is based on first-principles density functional theory (DFT) calculations, and the resulting energy-loss function exhibits a pronounced plasmon resonance at $\mathcal{O}(10)~\mathrm{eV}$~\cite{Liang:2024xcx,Guo:2024sqh,Gong:2026dte}.
As shown in Eq.~\ref{eq:materialcrosssection}, the energy-loss function enters the differential scattering cross section as the material response. When the SIDM kinematics allow access to the corresponding region of energy-momentum transfer phase space, collective electronic excitations enhance the expected ionization rate and thereby improve the sensitivity of semiconductor detectors to light SIDM.

For comparison with few-electron ionization data, we convert the deposited-energy spectrum into a charge-binned event rate. 
In Skipper-CCD detectors, the observable is the ionized charge $\mathcal Z$, defined as the number of electron--hole pairs produced by an energy deposition in silicon.
The event rate \(R_{\mathcal Z}\) is obtained by convolving the differential rate \(\dd R/\dd\omega\) with the secondary pair-creation probability distribution \(P(\mathcal Z|\omega)\):
\begin{equation}
 R_{\mathcal Z}
 =
 \int \dd \omega\,
 \frac{\dd R}{\dd \omega}
 P(\mathcal Z|\omega),
 \label{eq:charge_rate}
\end{equation}
where \(P(\mathcal Z|\omega)\) is the probability that an energy deposition \(\omega\) produces \(\mathcal Z\) electron--hole pairs. For the pair-creation probability distribution in silicon, we adopt the parametrization of Ref.~\cite{Ramanathan:2020fwm}, with band-gap energy \(E_{\rm gap}=1.16\ \text{eV}\) and mean energy per electron-hole pair \(\varepsilon=3.75\ \text{eV}\).
\begin{figure}[htbp]
 \centering
\includegraphics[width=\columnwidth]{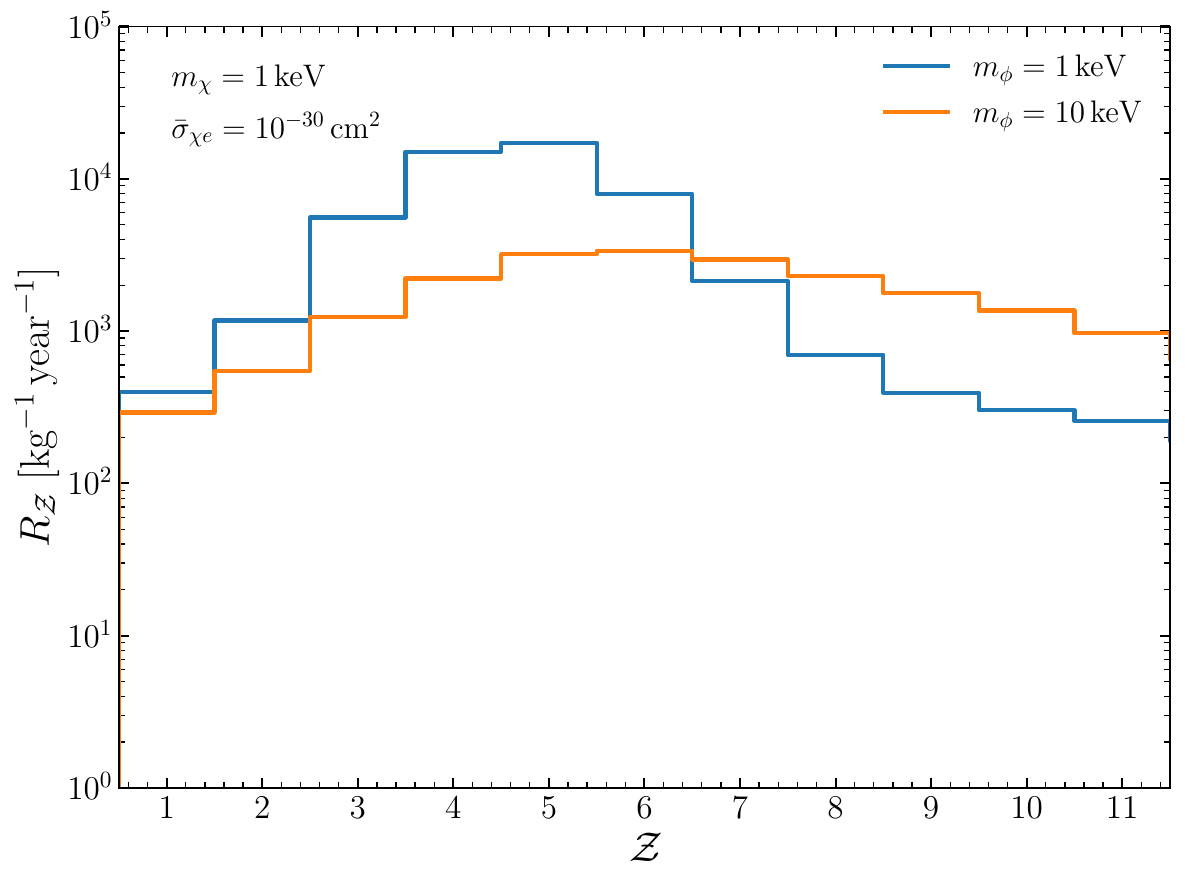}
 \caption{Event rate $R_{\mathcal Z}$ as a function of the number of electron-hole pairs $\mathcal Z$. We fix the DM mass at $\mchi=1~\mathrm{keV}$ and the reference DM--electron scattering cross section at $\sigmabar=10^{-30}~\mathrm{cm^2}$. Results are shown for two representative mediator masses, $\mphi=1~\mathrm{keV}$ and $\mphi=10~\mathrm{keV}$. For display purposes, 
 the curve corresponding to $\mphi=10~\mathrm{keV}$ is rescaled by a factor of $10^2$; all exclusion limits are derived using the unscaled event rates.
 }
 \label{fig:rate}
\end{figure}

Figure~\ref{fig:rate} shows the charge-binned event rate \(R_{\mathcal Z}\) for two representative mediator masses. The dependence on \(m_\phi\) originates from the mediator form factor: a lighter mediator enhances the small-momentum-transfer contribution, leading to a larger plasmon-induced response, while a heavier mediator suppresses this enhancement and shifts the spectral weight to higher \(\mathcal Z\). In Fig.~\ref{fig:rate}, the \(m_\phi = 10~\mathrm{keV}\) spectrum is scaled by \(10^2\) for visual clarity; all limits presented below are derived from the unscaled rates.
The light-mediator signal is concentrated mainly in $3\leq\mathcal Z\leq7$, whereas the heavier-mediator spectrum extends to larger $\mathcal Z$. We therefore use the 3--7 electron--hole pairs data, with the published selection efficiencies~\cite{SENSEI:2023zdf,DAMIC-M:2023gxo,DAMIC-M:2025luv}.

\section{Constraints from SENSEI and DAMIC-M}

We derive constraints by comparing the predicted charge-binned ionization
rates with public few-electron data from the SENSEI Skipper-CCD data set at
SNOLAB~\cite{SENSEI:2023zdf} and the DAMIC-M prototype-detector data set
collected at the Modane Underground Laboratory~\cite{DAMIC-M:2025luv}. The
observable in both analyses is the ionized charge \(\mathcal Z\), expressed
as the number of electron-hole pairs, which matches the event yields computed
in Sec.~III.

% For SENSEI, we use the \(\mathcal Z=3\)-7 charge bins with the reported
% bin-dependent exposures and selection efficiencies. For DAMIC-M, the
% published analysis classifies four-electron candidates by pixel-charge
% topology, whereas our signal calculation predicts only the total ionized
% charge. We therefore combine the \(\{31\}\), \(\{22\}\), and \(\{211\}\)
% topologies into a single \(\mathcal Z=4\) bin. This gives \(n_{4e}=1\) and
% \(b_{4e}=0.117\), obtained by summing the reported random-coincidence and
% radiogenic backgrounds. The signal yield is evaluated with an effective
% exposure of \(0.993~\mathrm{kg\,day}\), including the reported \(79\%\)
% four-electron pattern-selection efficiency. The DAMIC-M result should thus be interpreted as a charge-level recast rather than a topology-resolved
% likelihood. The charge-bin inputs are summarized in Table~\ref{tab:datasets}.

For SENSEI, we use the \(\mathcal Z=3\)--7 charge bins together with the
reported candidate counts, background expectations, and bin-dependent
effective exposures. The DAMIC-M analysis identifies candidate events as charge patterns formed by
two or three adjacent pixels. Since our calculation predicts only the total
ionized charge $\mathcal{Z}$, we combine the patterns corresponding to the
same total charge. Specifically, the $\{21\}$ and $\{111\}$ patterns are used
for $\mathcal{Z}=3$, while the $\{31\}$, $\{22\}$, and $\{211\}$ patterns
are used for $\mathcal{Z}=4$. The observed counts and background expectations
are obtained by summing the corresponding pattern-level values reported by
DAMIC-M. The event-selection criteria were defined using the first seven days
of data, denoted D1, and subsequently applied to the remaining data set,
denoted D2, in a blind analysis. Only the D2 data set is used for the DM
search, with an integrated exposure of $1.257~\mathrm{kg\,day}$. The reported
efficiencies for three- and four-electron deposits to be selected as one of
the considered charge patterns are approximately $65\%$ and $79\%$,
respectively. The corresponding effective exposures, together with the
observed counts and background expectations, are summarized in
Table~\ref{tab:datasets}. Since
the full topology-response probabilities are
not publicly available, our DAMIC-M result is an approximate total charge recast rather than a reproduction of the official pattern-level likelihood. Further details are
provided in Appendix~\ref{app:damic_recast}.

\begin{table}[htbp]
\caption{Data from the SENSEI and DAMIC-M experiments used in this work. $\mathcal Z$ denotes the ionized charge.}
\centering
\begin{tabular}{c|cccc}
\hline
Experiment & Charge bin & Events & Background & Exposure \\
\hline
\multirow{5}{*}{SENSEI}
 & $\mathcal Z=3$ & $4$ & $0.07$ & $57.71~\mathrm{g\,day}$ \\
\cline{2-5}
 & $\mathcal Z=4$ & $0$ & $0$ & $63.03~\mathrm{g\,day}$ \\
\cline{2-5}
 & $\mathcal Z=5$ & $0$ & $0$ & $65.56~\mathrm{g\,day}$ \\
\cline{2-5}
 & $\mathcal Z=6$ & $0$ & $0$ & $67.31~\mathrm{g\,day}$ \\
\cline{2-5}
 & $\mathcal Z=7$ & $0$ & $0$ & $68.53~\mathrm{g\,day}$ \\
\hline
\multirow{2}{*}{DAMIC-M}
 & $\mathcal Z=3$ & $0$ & $0.208$ & $0.817~\mathrm{kg\,day}$ \\
\cline{2-5}
 & $\mathcal Z=4$ & $1$ & $0.117$ & $0.993~\mathrm{kg\,day}$ \\
\hline
\end{tabular}
\label{tab:datasets}
\end{table}

For each benchmark value of $\sigmabar$, we scan the
$(\mchi,\mphi)$ plane and calculate the predicted signal yield
$s_i$ in charge bin $i$. Since the same
DM--electron interaction governs both cosmic-ray upscattering and
the subsequent scattering in silicon, the signal yield scales as
$s_i\propto\sigmabar^2$. Assuming independent Poisson fluctuations
among the charge bins, the likelihood is
\begin{equation}
 \mathcal L
 =
 \prod_i
 \frac{(s_i+b_i)^{n_i}}{n_i!}
 \exp[-(s_i+b_i)],
 \label{eq:likelihood}
\end{equation}
% \begin{equation}
%  \begin{aligned}
%  \mathcal L(\mchi,\mphi,\sigmabar)
%  &=
%  \prod_i
%  \frac{
%  \left(s_i(\mchi,\mphi,\sigmabar)+b_i\right)^{n_i}
%  }{n_i!} \\
%  &\quad\times
%  \exp\!\left[-s_i(\mchi,\mphi,\sigmabar)-b_i\right],
%  \end{aligned}
%  \label{eq:likelihood}
% \end{equation}
where $n_i$ and $b_i$ are the observed and expected background event
counts in charge bin $i$, respectively. The background expectations are fixed to their reported central values in the present analysis.
At each point in the $(\mchi,\mphi)$ plane, the corresponding
signal-plus-background hypothesis is tested using the one-sided
likelihood-ratio statistic %($\widetilde q_{\sigmabar}$) 
and the $\mathrm{CL}_s$ prescription, as implemented in
\textsc{pyhf}~\cite{Cowan:2010js,Heinrich:2021gyp}.
We exclude a parameter point at the 90\% confidence level when $\mathrm{CL}_s<0.1$.

\begin{figure}[!t]
 \centering
 \includegraphics[width=\columnwidth]{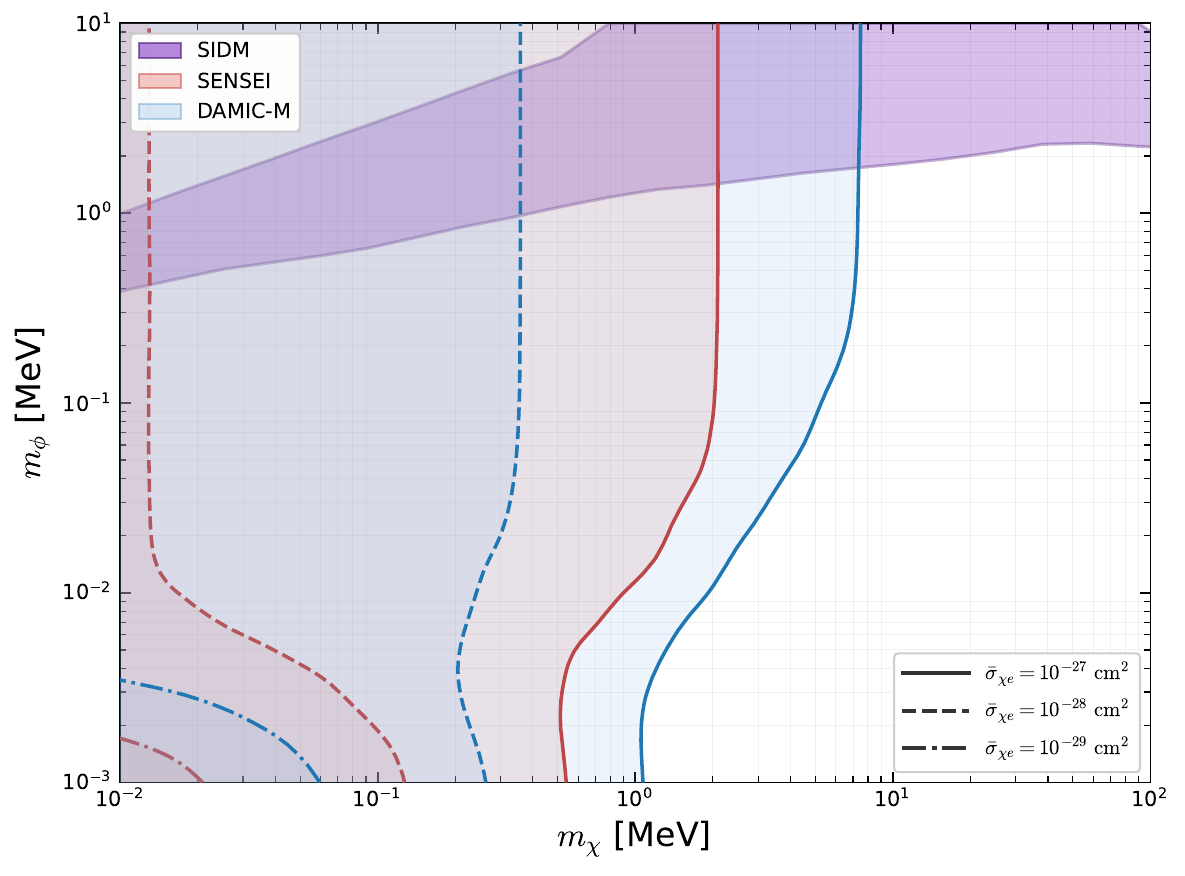}
 \caption{Constraints on the \((\mchi,\mphi)\) plane from SIDM--electron
scattering in silicon at 90\% C.L. Red regions are obtained from the SENSEI experiment, while the blue regions show the constraints obtained from the
DAMIC-M experiment. Line styles denote
\(\sigmabar=10^{-27}\,\mathrm{cm^2}\) (solid),
\(10^{-28}\,\mathrm{cm^2}\) (dashed), and
\(10^{-29}\,\mathrm{cm^2}\) (dash-dotted). The purple band denotes the SIDM target region satisfying
$0.1<\langle\sigma_T\rangle/m_\chi
<10~\mathrm{cm^2/g}$, obtained by scanning
$10^{-3}<\alpha_\chi<1$. 
The parameter space is additionally restricted to $g_e<1$.
The largest benchmark value, \(\sigmabar=10^{-27}~\mathrm{cm^2}\), is chosen to avoid entering
the regime in which attenuation of the accelerated SIDM flux in
the atmosphere or overburden may become important~\cite{Chen:2021ifo,DAMIC-M:2025luv}.
}
 \label{fig:limits}
\end{figure}

Figure~\ref{fig:limits} shows the resulting 90\% C.L. exclusion contours in the \((\mchi,\mphi)\) plane for representative values of \(\sigmabar\). The purple band denotes the SIDM target region defined in Eq.~\eqref{eq:sidmtarget}. The red regions are obtained from the SENSEI experiment, while the blue regions correspond to the DAMIC-M experiment.
Their overlap with the SIDM target band shows that few-electron silicon ionization data probe part of the light-mediator SIDM parameter space favored by galactic small-scale anomalies.

The excluded regions expand with increasing \(\sigmabar\), as this parameter controls both the cosmic-ray upscattering rate and the subsequent ionization yield. The dependence on \(\mphi\) is nontrivial, as this mass governs the mediator range, the momentum-transfer dependence of the scattering, and the electronic response of the silicon target.

\section{Conclusions}
In this work, we investigate the sensitivity of silicon Skipper-CCD detectors to light SIDM accelerated by high-energy galactic cosmic ray electrons. Considering a light scalar mediator coupled to both DM and electrons, we account for both velocity-dependent DM self-interactions and DM-electron scattering in silicon. Cosmic ray upscattering produces a semi-relativistic SIDM component that can access the energy-momentum-transfer region of electronic collective excitations in silicon. Using the silicon energy-loss function $\mathrm{Im}[-\epsilon^{-1}(q,\omega)]$, we calculate the resulting electronic excitation event rate and convert it into charge-binned ionization rates.

Using public SENSEI and DAMIC-M ionization data, we derived 90\% C.L. exclusion limits in the $(m_\chi,m_\phi)$ plane for benchmark values of the reference cross section $\bar{\sigma}_{\chi e}$. Our constraints can cover a portion of the light SIDM parameter space satisfying $0.1 < \langle\sigma_T\rangle/m_\chi < 10~\mathrm{cm^2/g}$. These results show that plasmon excitations in silicon provide sensitivity to light SIDM through few-electron ionization signals, thereby offering a complementary laboratory probe of scenarios in which the mediator responsible for DM self-interactions also couples
to electrons.
Future sensitivity may be improved through larger low-background
Skipper-CCD exposures, higher event-selection efficiencies, and more
precise calibration of the detector response at low deposited energies.
The same framework may also be extended to other semiconductor targets,
such as germanium and gallium arsenide, which could probe complementary
regions of the light SIDM parameter space.

\section*{Acknowledgments}
This work was supported by the National Natural Science Foundation of China
(NSFC) under Grant No.~12275134 and by the Talent Program of Chengdu
Technological University under Grant No.~2024RC031.
\appendix

\section{Self-interaction cross section}
\label{app:selfinteraction}
In the semiclassical regime ($\kappa\gtrsim1$), we use the analytic expression for an attractive
potential derived in Ref.~\cite{Colquhoun:2020adl}:
\begin{equation}
 \sigma_T^{\rm att}=\frac{\pi}{\mphi^2}
 \begin{cases}
 2\beta^2\zeta_{1/2}(\kappa,\beta), & \beta\leq0.2,\\
 2\beta^2\zeta_{1/2}(\kappa,\beta)
 e^{0.64(\beta-0.2)}, & 0.2<\beta\leq1,\\
 4.7\ln(\beta+0.82), & 1<\beta<50,\\
 2\ln\beta\,(\ln\ln\beta+1), & \beta\geq50,
 \end{cases}
 \label{eq:semiclassical}
\end{equation}
where
\begin{align}
 \zeta_n(\kappa,\beta)
 &=
 \frac{\max(n,\beta\kappa)^2-n^2}{2\kappa^2\beta^2}
 +\eta\!\left(\frac{\max(n,\beta\kappa)}{\kappa}\right), \nonumber\\
 \eta(x)&=x^2\left[-K_1^2(x)+K_0(x)K_2(x)\right],
 \label{eq:zeta}
\end{align}
and $K_n$ denotes a modified Bessel function of the second kind.

For $\kappa\lesssim0.4$, scattering is dominated by the $s$-wave. In this
regime, we approximate the Yukawa potential by the analytically solvable
Hulth\'{e}n potential, which gives
\begin{equation}
 \sigma_T^{\rm Hulth\acute en}
 =\frac{16\pi}{\mchi^2v^2}\sin^2\delta_0,
 \label{eq:hulthen}
\end{equation}
where $\delta_0$ is the corresponding $s$-wave phase shift~\cite{Tulin:2013teo}. 
It is given by
\begin{equation}
\label{eq:delta0_lambda}
\begin{aligned}
    \delta_0 &= \arg \left[ 
        i \, \frac{\Gamma(\lambda_+ + \lambda_- - 2)}{\Gamma(\lambda_+) \, \Gamma(\lambda_-)}
    \right], \\
    \lambda_{\pm} &= 1 \pm \frac{i \kappa}{1.6} \left( 1 + \sqrt{1 - 3.2\beta} \right)\,.
\end{aligned}
\end{equation}
In the intermediate region
$0.4<\kappa<1$, we interpolate smoothly between the quantum and
semiclassical expressions:
\begin{equation}
 \sigma_T=
 \frac{1-\kappa}{0.6}\sigma_T^{\rm Hulth\acute en}
 +\frac{\kappa-0.4}{0.6}\sigma_T^{\rm att}.
 \label{eq:interpolation}
\end{equation}
This interpolation provides a continuous semi-analytic cross section that
smoothly connects the two regimes.
\section{Charge-bin inputs for the DAMIC-M recast}
\label{app:damic_recast}

The DAMIC-M search reports its low-charge candidates in terms of
pixel-charge patterns formed by two or three adjacent pixels. Our signal
calculation, by contrast, predicts the yield as a function of the total
ionized charge. To use the published pattern-level information in a
charge-binned likelihood, we combine patterns with the same total
reconstructed charge,
\begin{equation}
 {\cal P}_{3e^-}=\{\{21\},\{111\}\},
 \qquad
 {\cal P}_{4e^-}=\{\{31\},\{22\},\{211\}\}.
\end{equation}
For an aggregated charge bin \(\mathcal Z\), the observed count and the
central background expectation are then
\begin{equation}
 n_{\mathcal Z}
 =
 \sum_{p\in{\cal P}_{\mathcal Z}} D_p,
 \qquad
 b_{\mathcal Z}
 =
 \sum_{p\in{\cal P}_{\mathcal Z}}
 \left(B_p^{\rm rc}+B_p^{\rm rad}\right),
 \label{eq:damic_bin_aggregation}
\end{equation}
where \(D_p\) is the number of candidates in pattern \(p\), and
\(B_p^{\rm rc}\) and \(B_p^{\rm rad}\) denote the random-coincidence
and radiogenic background components reported for pattern \(p\). This
corresponds to evaluating the DAMIC-M radiogenic-background normalization at
its central value. 
%The full experimental likelihood instead treats the
%radiogenic normalization as a nuisance parameter through
%\(B_p=B_p^{\rm rc}+\theta B_p^{\rm rad}\).
The pattern-level inputs used in the aggregation are listed in
Table~\ref{tab:damic_patterns}.

\begin{table}[htbp]
\centering
\begin{tabular}{|c|c|c|c|}
\hline
 & \multicolumn{3}{c|}{Pattern $p$} \\
\cline{2-4}
 & $\{11\}$ & $\{21\}$ & $\{111\}$ \\
\hline
$D_p$
& $144$ & $0$ & $0$ \\
$B_p^{\rm rc}$
& $141.4$ & $0.111$ & $0.042$ \\
$B_p^{\rm rad}$
& $0.039$ & $0.039$ & $0.016$ \\
\hline
 & $\{31\}$ & $\{22\}$ & $\{211\}$ \\
\hline
$D_p$
& $1$ & $0$ & $0$ \\
$B_p^{\rm rc}$
& $0.019$ & $2.5\times10^{-5}$ & $5.8\times10^{-5}$ \\
$B_p^{\rm rad}$
& $0.052$ & $0.011$ & $0.035$ \\
\hline
\end{tabular}
\caption{Candidate counts \(D_p\) in the D2 data set and expected background counts from random coincidences, \(B_p^{\rm rc}\), and radioactive decays, \(B_p^{\rm rad}\).}
\label{tab:damic_patterns}
\end{table}

The resulting three-electron bin has no observed candidates and a total background:
\begin{align}
 n_{3e^-} &= 0, \nonumber\\
 b_{3e^-}
 &= (0.111+0.039)+(0.042+0.016)
 =0.208.
 \label{eq:damic_three_electron_input}
\end{align}
For the four-electron bin, the single observed event comes from the
\(\{31\}\) pattern, while the \(\{22\}\) and \(\{211\}\) patterns contain no
candidates. We therefore use
\begin{align}
 n_{4e^-} =& 1, \nonumber\\
 b_{4e^-}
 =&
 (0.019+0.052)
 +(2.5\times10^{-5}+0.011) \nonumber\\
 &+(5.8\times10^{-5}+0.035)
 \simeq 0.117.
 \label{eq:damic_four_electron_input}
\end{align}

% The signal expectation in a charge bin is normalized with an effective
% exposure,
% \begin{equation}
%  s_{\mathcal Z}
%  =
%  R_{\mathcal Z}\,
%  {\cal E}_{\mathcal Z}^{\rm eff},
%  \qquad
%  {\cal E}_{\mathcal Z}^{\rm eff}
%  =
%  {\cal E}_{\rm D2}\epsilon_{\mathcal Z},
%  \label{eq:damic_effective_exposure}
% \end{equation}
% where \(R_{\mathcal Z}\) is the predicted rate in charge bin \(\mathcal Z\),
% \({\cal E}_{\rm D2}=1.257~\mathrm{kg\,day}\) is the D2 exposure, and
% \(\epsilon_{\mathcal Z}\) is the DAMIC-M pattern-selection efficiency for a
% deposit with \(\mathcal Z\) electron-hole pairs. Using
% \(\epsilon_{3e^-}=0.65\) and \(\epsilon_{4e}=0.79\), corresponding to the
% reported efficiencies for three- and four-electron deposits to pass the
% accepted-pattern selection, gives
% \begin{equation}
%  {\cal E}_{3e^-}^{\rm eff}
%  =
%  0.817~\mathrm{kg\,day},
%  \qquad
%  {\cal E}_{4e}^{\rm eff}
%  =
%  0.993~\mathrm{kg\,day}.
% \end{equation}

% The two-electron \(\{11\}\) category is not included in this recast. It is
% background dominated, and a consistent treatment would require the
% pattern-resolved likelihood and its associated nuisance parameters. Our
% charge-bin construction also neglects off-diagonal migrations between the
% generated ionization charge and the reconstructed pixel pattern. The result
% should therefore be interpreted as an approximate charge-binned
% reinterpretation of the DAMIC-M data, rather than a reproduction of the
% official pattern-level likelihood.
\bibliography{refs}

\end{document}